\documentclass{article}

\usepackage{PRIMEarxiv}

\usepackage[utf8]{inputenc} 
\usepackage[T1]{fontenc}    
\usepackage{hyperref}       
\usepackage{url}            
\usepackage{booktabs}       
\usepackage{amsfonts}       
\usepackage{nicefrac}       
\usepackage{microtype}      
\usepackage{lipsum}
\usepackage{fancyhdr}       
\usepackage{graphicx}       
\usepackage{braket}    

\graphicspath{{figures/}} 

\title{Improvement in Variational Quantum Algorithms
by Measurement Simplification
}

\author{
  Jaehoon Hahm \\
  Department of Physics and Astronomy \\
  Seoul National University \\
  \texttt{judol1snu@ac.kr} \\
   \And
  Hayeon Kim, Young June Park \\
  Department of Electrical and Computer Engineering \\
  Seoul National University \\
  \texttt{\{khy5630, ypark\}@snu.ac.kr} \\
}

\begin{document}
\maketitle

\begin{abstract}

Variational Quantum Algorithms (VQAs) are expected to be promising algorithms with quantum advantages that can be run at quantum computers in the close future. In this work, we review simple rules in basic quantum circuits, and propose a simplification method, Measurement Simplification, that simplifies the expression for the measurement of quantum circuit. By the Measurement Simplification, we simplified the specific VQA’s result expression and obtained large improvements in calculation time and required memory size. Here we applied Measurement Simplification to Variational Quantum Linear Solver (VQLS), Variational Quantum Eigensolver (VQE) and other Quantum Machine Learning Algorithms to show an example of speedup in the calculation time and required memory size.

\end{abstract}


\section{Introduction}

After the discovery of Shor's algorithm and Grover's search algorithm, there has been many researches covering the concept of quantum advantage, which insists quantum computers will exhibit specific advantages over classical computers. Google named the advantage as "Quantum Supremacy"\cite{refbib1} and explains for specific problems, quantum computers can surpass classical computer in computation time and required memory capacity. However, complex quantum algorithms such as Shor's algorithm requires number of qubits and gate fidelity exponentially more than currently we have, and therefore investigating executable algorithms that show quantum advantage even with noisy and few qubits have been arised as an important question in the NISQ (Noisy Intermediate-Scale Quantum) era\cite{refbib2}.

Among them, VQAs (Variational Quantum Algorithms)\cite{refbib3}  have been remarked as efficient algorithms that can been executed in NISQ devices with low limitation. VQA is a hybrid quantum algorithm that utilizes classical optimizer and Variational Quantum Circuit (VQC), it first measures a state's probability after quantum circuit, and passes the result to classical optimizer. Then, the optimizer outputs a feedback to the quantum circuit and the repetition of this overall process optimizes the quantum circuit in a direction that outputs a correct measurement probability. 

For example, Quantum Deep Reinforcement Learning (QDRL)\cite{refbib4}  is a VQA version of Deep Reinforcement Learning, which utilizes the memory advantage exhibited by neural network learning, and there have been reports that QDRL can perform same reinforcement learning but with exponentially smaller memory  than the classical counterpart. Also, Variational Quantum Eigensolver (VQE) \cite{refbib5} \cite{refbib6} is a prominent quantum algorithm that can be utilized to calculate molecule's ground state energy and optimize asset portfolio. Researches of Quantum Machine Learning (QML) \cite{refbib8}  that  exploit Quantum Kernel, Quantum Support Vector Machine (QSVM)\cite{refbib7} for image classification are actively progressing. Furthermore, papers are being reported to solve max-cut problem, which is a graph optimization problem, and knapsack problem, which is a combinatorial optimization problem that can be exploited in battery revenue optimization, which show VQA is a prominent solving method for various optimization problems.

On the other hand, certain quantum algorithms does not use entire information of output state vector, but only use a specific part of the full information, such as expectation value with respect to specific basis, or an amplitude of $\ket{0^n}$, or the probability that the most significant bit (MSB) is measured as 0. Specific VQAs only use partial information of the output state in the repeated feedback loop and progresses the optimization. For example, at the quantum chemistry simulation using VQE, it measures the Hamiltonian's expectation value in order to calculate a molecule's ground state energy. This is equivalent to measure the transition amplitude $\bra{0^n}U^\dagger{} HU\ket{0^n}$ , where $U$ is an unitary operator that satisfies $\ket{\psi} = U\ket{0^n}$, and hence equivalent to using the coefficient of the first basis of the output state vector which corresponds to the amplitude of $\ket{0^n}$. Therefore, it is possible to execute equivalent algorithm with only partial information, and if it is possible to simplify the output measurement expression, there exists benefit in computation time and required memory capacity. In this research, we named the process of simplifying the output measurement expression as "Measurement Simplification" and showed the simplification is valid for various important VQAs.

One of the important VQAs, Variational Quantum Linear Solver (VQLS) \cite{refbib10} is an algorithm for finding solution of linear equations, which is A$\vec{x}=\vec{b}$ where A, $\vec{b}$ are given constant matrix and vector respectively, and $\vec{x}$ is the desired solution. Harrow-Hassidim-Lloyd or HHL Algorithm \cite{refbib12} is an algorithm that exhibits exponential speedup in specific linear equation problems, and VQLS is its VQA version. \cite{refbib10} reports VQLS shows similar quantum advantages compared to HHL's algorithm. In this research, we showed VQLS gains substantially large advantage in computation time and required memory capacity after Measurement Simplification.

\section{Measurement Simplification in Basic Quantum Circuits}
\label{sec:ms_basic}

\begin{figure}
	\includegraphics[width=0.6\textwidth]{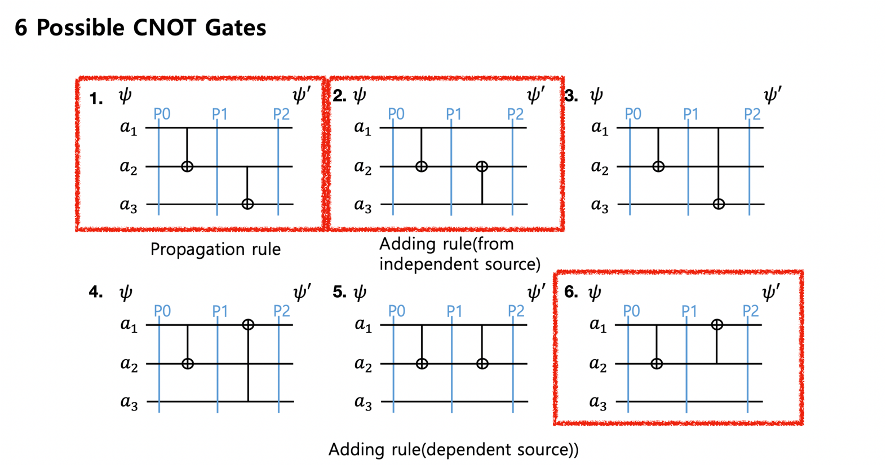}
	\centering
	\caption{6 possible CNOT gates combination for n=3 qubits case. There exists simple rules for simplifying the output measurement expression.}
	\label{fig:fig1}
\end{figure}

\begin{figure}
	\includegraphics[width=0.6\textwidth]{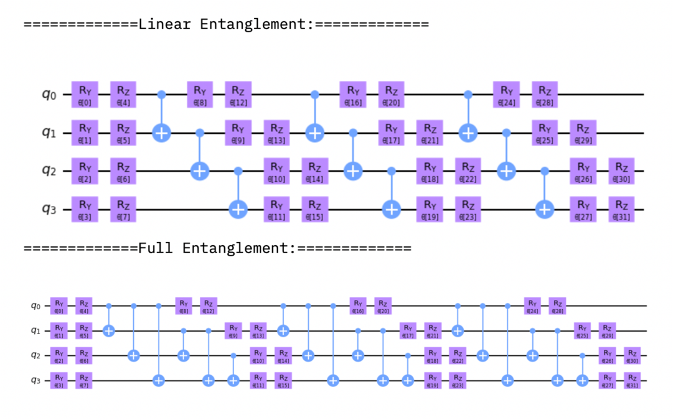}
	\centering
	\caption{Parametrized quantum circuit for calculating $H_2$ molecule's ground state energy by VQE method.}
	\label{fig:fig2}
\end{figure}

\begin{figure}
	\includegraphics[width=0.6\textwidth]{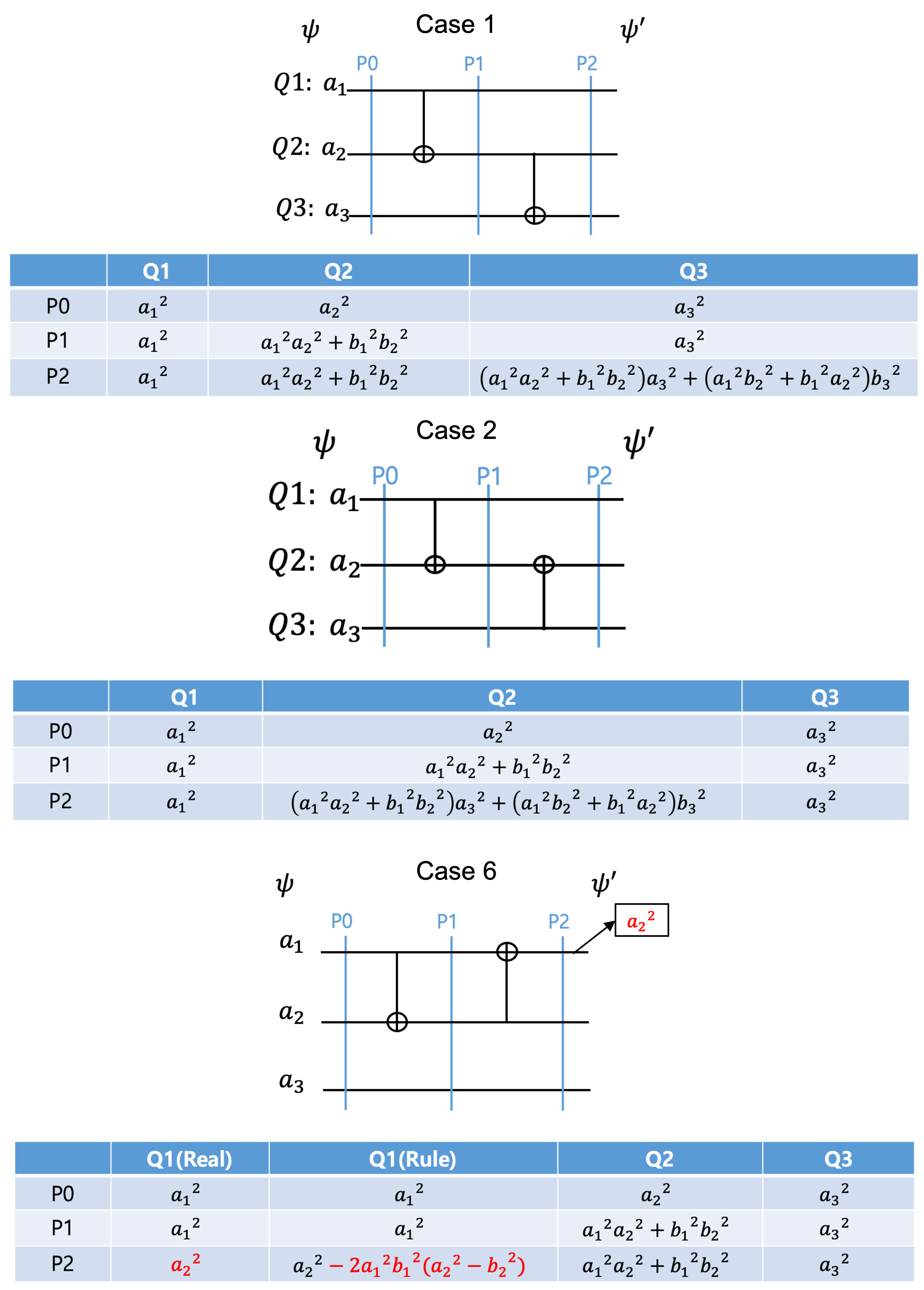}
	\centering
	\caption{Simplification rules for CNOT circuits}
	\label{fig:result1}
\end{figure}

For CNOT gates applied to 3-qubits quantum circuit, the cases can be classified into 6 cases as Figure \ref{fig:fig1}. Figure \ref{fig:fig2} is a parametrized circuit that calculates $H_2$ molecule's ground state energy using VQE method. It can be seen that Figure \ref{fig:fig1}'s CNOT structure is similar to that of Figure \ref{fig:fig2}. If we focus on the probability of the 1st qubit to be measured as 0 at the 1st, 2nd, and 6th circuit of Figure \ref{fig:fig1}, simple rules can be found which is described at Figure \ref{fig:result1}. Q1, Q2, Q3 respectively denotes the probability for 1st, 2nd, 3rd qubit to be measured as $\ket{0}$ and P0, P1, P2 denotes the phase or stage which the state is present in the circuit. We supposed the input state is a separable state and denoted the probability of $i$th qubit to be measured as 0 as $a_i^2$, and 1 as $b_i^2$.

For Case 1, the effect of CNOT can be described as transiting the value of control Q1 to target Q2, in other words, the probability of the target to measured as 0 transits as $a_2^2 \to a_1^2a_2^2 + b_1^2b_2^2$. Then, if we once again apply CNOT to control Q2, target Q3, there exhibits the same effect. This implies that the expression of Q1, the probability of the 1st qubit to be measured as 0, is factorizable and the expression can be simplified using the identity $a_i^2 + b_i^2 = 1$. We named this process as "Measurement Simplification" or shortly simplification and the simplification transformation as "rule". Case 2 is almost the same circuit with case 1 but cnot target applied to identical qubit. The simplification shows the composition of rules is also available for the specific cases.

For Case 6, it is the case that the target and control being reversed at the 2nd CNOT. It can be seen that the rule is successful until P1 but fails at P2. In particular, if we explicitly simplify, we obtain $Q_1 = a_2^2$. This implies for the case 6 circuit structure, if we only require partial information such as $Q_1$, we may exploit the Measurement Simplification and reduce the expression size, which will lead to advance in computation speed. If we require the information of $Q_2$, the advantage is smaller than the case for $Q_1$ but the expression size still decreases substantially after simplification. Moreover, the CNOT structure as in case 6 can be interpreted as basis permutation of the state vector, and since it is the 3-cycle permutation, the circuit is equivalent to identity for every 3 repetitions of the circuit.

\begin{figure}
	\includegraphics[width=0.6\textwidth]{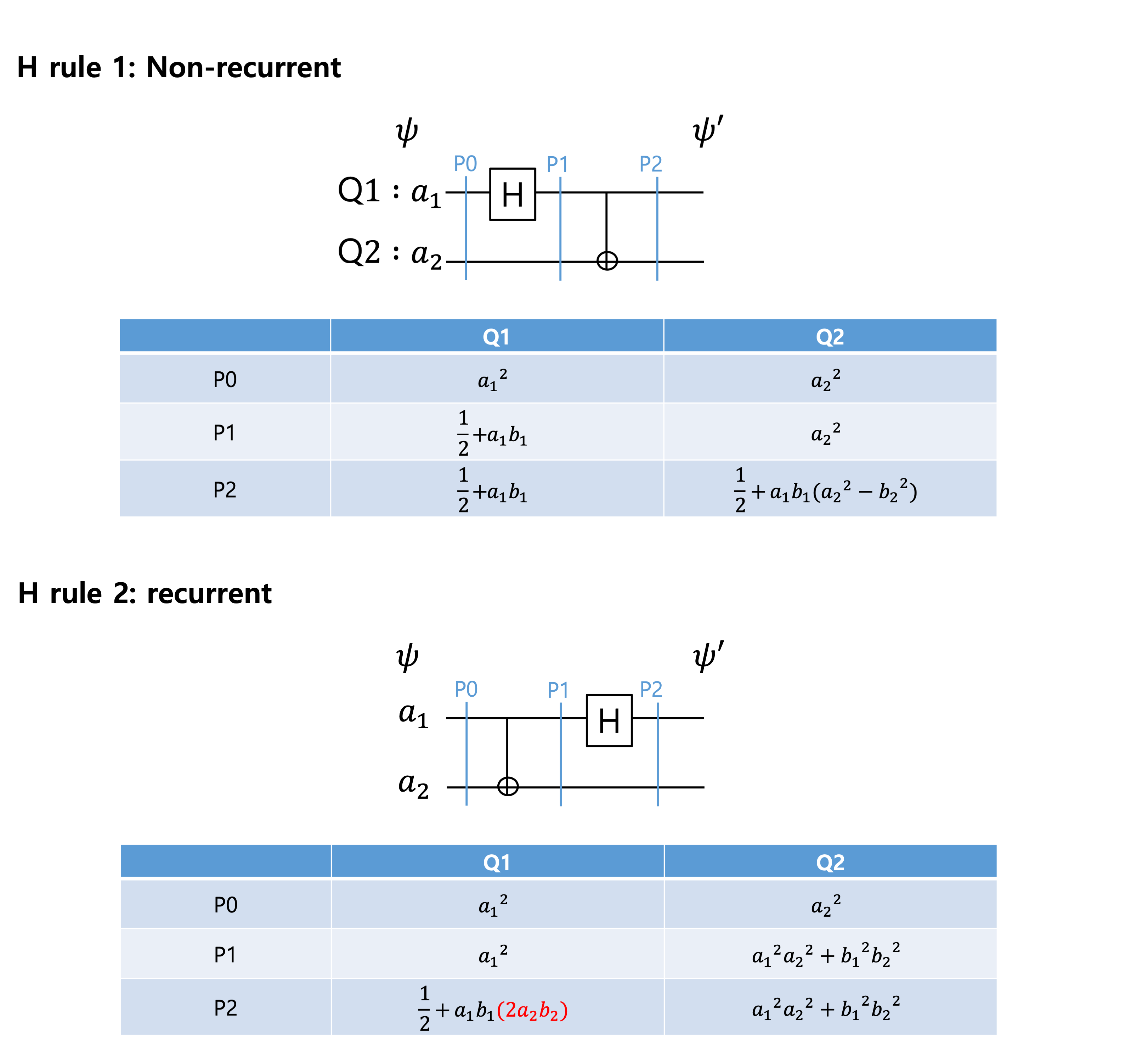}
	\centering
	\caption{Simplification rules for CNOT circuits}
	\label{fig:result2}
\end{figure}

Figure \ref{fig:result2} illustrates the simple circuits composed of Hadamard gate and CNOT gate. The circuits can be interpreted as circuit that generates and disentangles the Bell state respectively. For case 1, Hadamard gate exhibits the rule of $a_1^2 \to 1/2 + a_1b_1$. After CNOT, it showed the effect of $a_2^2 \to 1/2 + a_1b_1(a_2^2-b_2^2)$ which was inconsistent with the CNOT rule we have found previously. This implies that simplification rule is dependent whether the input state is separable or entangled and we need to consider complex simplification rules if we are considering a general input state.

However, there still exists the decrease of expression size after simplification, and this advatange is utilizable in algorithms that uses only partial information of the full state vector, such as VQA. In other words, by giving up the amplitude information of unimportant bases, one may decreases the expression size and gain speedup in computation time, and advantage in required memory capacity in compensation. We named this process as Measurement Simplification, and showed that exploiting this method, large speedup is possible in various VQAs, such as Quantum Deep Reinforcement Learning (QDRL), Variational Quantum Linear Solver (VQLS) and Quantum Chemistry problems.

\section{Measurement Simplification in Variational Quantum Algorithm (VQA)}
\label{sec:ms_vqa}

\subsection{Application in Quantum Deep Reinforcement Learning (QDRL)}

\begin{figure}
	\includegraphics[width=0.8\textwidth]{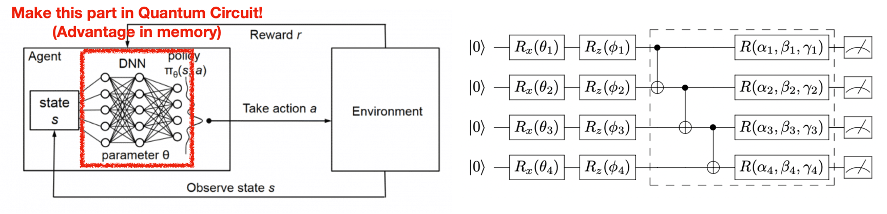}
	\centering
	\caption{Left Diagram explains the schematic of QDRL. Right diagram shows the VQC exploited in QDRL. }
	\label{fig:qdrl}
\end{figure}

Quantum Deep Reinforcement Learning (QDRL) is a quantum computing version of Deep Reinforcement Learning (DRL), which utilizes the neural network in Reinforcement Learning; it replaces the neural network part in DRL with Variational Quantum Circuit (VQC)\cite{refbib4}. The $R_x$, $R_z$ gates at the first part of VQC are intializer that is independent to parameter learning. The $\alpha$, $\beta$, $\gamma$ in the dashed box is the parameters that needs to be learned and the componenets in the dashed box can be repeated.

\begin{figure}
	\includegraphics[width=0.8\textwidth]{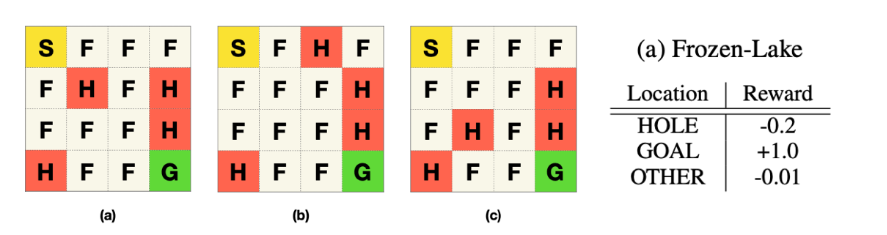}
	\centering
	\caption{Diagram explaining "Frozen-Lake" problem's environment.}
	\label{fig:lake}
\end{figure}

Figure \ref{fig:lake} is the diagram explaining the "Frozen-Lake" problem's environment that is solved at \cite{refbib4} using QDRL. It is the problem finding the shortest path that starts from S, while there exists reward at G and penalty at H. \cite{refbib4} reports that exploiting QDRL, it was able to attain comparable learning result with requiring exponentially smaller amount of memory. Specifically, if the number of qubits to define the problem is denoted as $n$, the number of parameters that needs to be stored and optimized is for classical reinforcement learning, $O(n^3)$, for DRL, $O(n^2)$, for QDRL, $O(n)$. Therefore, further optimizing the QDRL method would lead to a meaningful speedup in reinforcement learning. 

Among the VQCs, we fixed the number of qubits $n=2$, and investigated whether Measurement Simplification is possible. In other words, we verified whether the expression of probability for the 1st qubit to measured as 0 can be simplified. To realize the method of Measurement Simplification, we first calculated the matrix equivalent to the given quantum circuit using Mathematica, then producted appropriate basis to obtain the desired measurement expression, and used simplify function implemented in Mathematica to obtain the simplified expression.

\begin{figure}
	\includegraphics[width=0.8\textwidth]{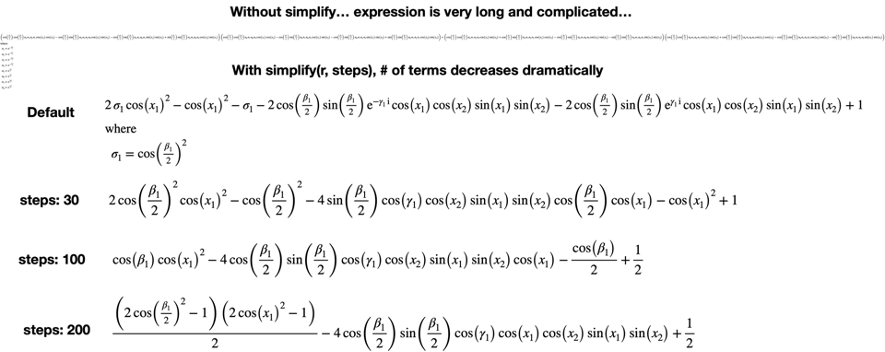}
	\centering
	\caption{After simplification, the number of terms in the measurement expression decreases dramatically.}
	\label{fig:simplify}
\end{figure}

Figure \ref{fig:simplify} shows the result of simplification, increasing the number of steps of simplify function, or simplification intensity. Before simplification, $Q_1$ had a very long expression, but as simplification intesntiy increases, the number of terms in expression decreases dramatically. Using Mathematica, we were able to confirm that the leafcount of the expression decreased $2273 \to 757$ 3 times smaller where leafcount is the number of indivisible subexpressions of the given expression. Hence, leafcount can be one of the metrics to measusre the expression's total size, and it can be sufficiently thought to be proportional to the calculation time of the expression evaluation.

Figure \ref{fig:simplify2} shows the simplification result for various VQCs including the circuit used for DQRL. We defined the improvement factor as the ratio of the leafcount of initial expression to the leafcount of expression after simplification. The improvement factor values were different for each circuits, but showed about 3 times improvement in average. In other words, utilizing the Measurement Simplification, one can obtain the exact same evaluation with 3 times shorter expression, and expect to gain advantage in computation time and required memory capacity similar to the ratio.

\begin{figure}
	\includegraphics[width=0.6\textwidth]{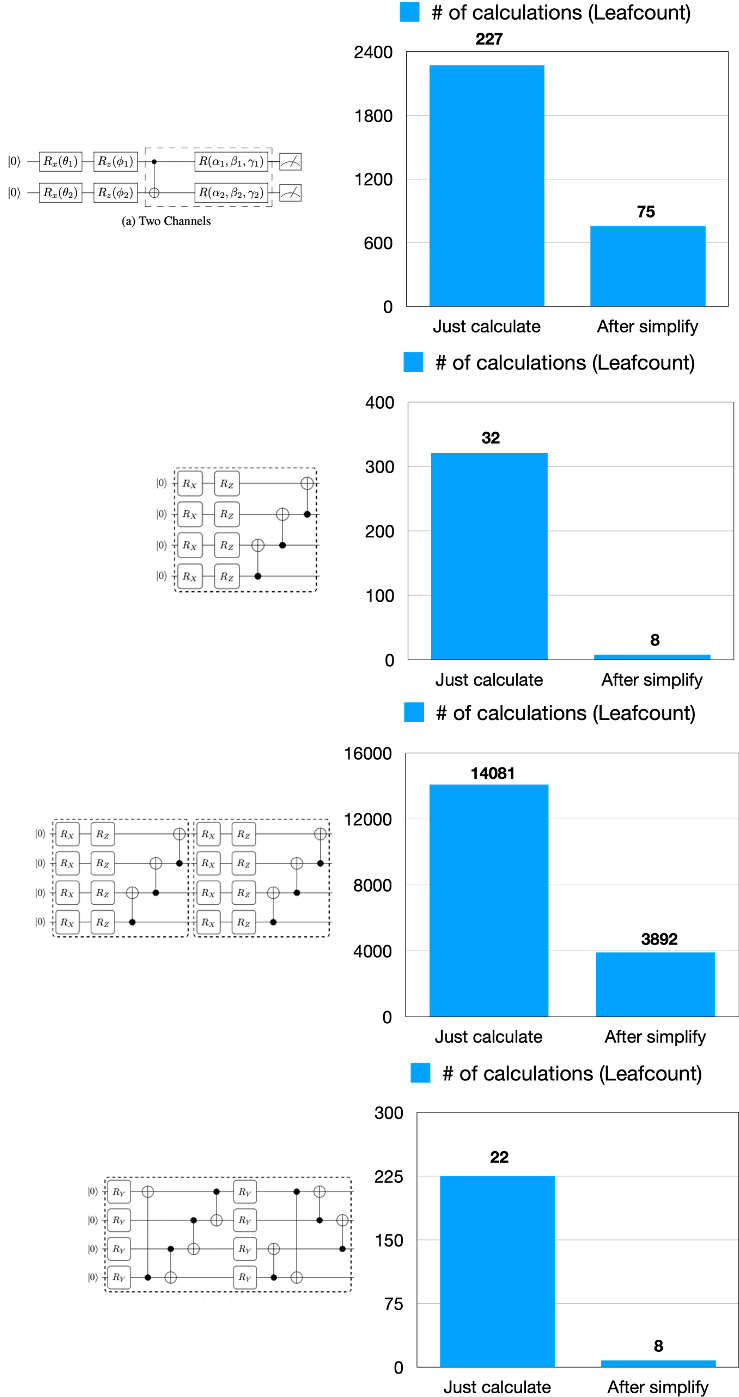}
	\centering
	\caption{For various VQCs, size of expression (leafcount) decreases after Measurement Simplification.}
	\label{fig:simplify2}
\end{figure}

\subsection{Application in Variational Quantum Linear Solver (VQLS)}

\begin{figure}
	\includegraphics[width=0.6\textwidth]{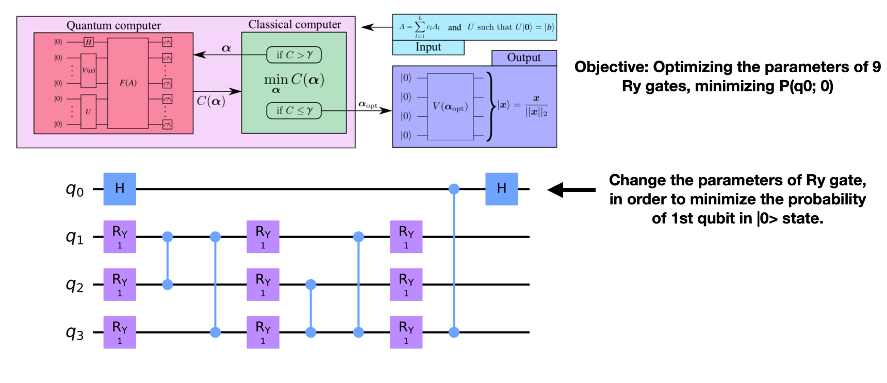}
	\centering
	\caption{Diagram explaining Variational Quantum Linear Solver (VQLS). Below circuit is the VQC that is utilized at VQLS.}
	\label{fig:vqls}
\end{figure}

Variational Quantum Linear Solver (VQLS) is a algorithm for solving linear equation, $A\vec{x} = \vec{b}$. HHL algorithm is a quantum algorithm that for specific matrices, exhibits exponential speedup over classical method and VQLS is a VQA version linear solver algorithm that has similar complexity compared to the HHL algorithm. Figure \ref{fig:vqls} schematically expalains VQLS; quantum circuit is composed of a parametrized ansatz and the cost function is encoded in the probability of the 1st qubit to be measured as 0. There are specifically two types of quantum circuits that calculates inner product of vectors, which is implemented by Hadamard test method. When the circuits output the cost function, this is given to the classical optimizer, and the optimizer again gives a feedback to the parametrized ansatz based on the input cost function, optimizing the parameter in the dirction to minimizing the cost function. In other words, for the parametrized quantum circuit $U(\theta)$, optimizing the parameters is equivalent to finding parameter $\vec{\theta}$, that makes $\ket{\psi}$ = $U(\theta)\ket{0}$ closest to the solution $\ket{x}$ of $A\vec{x} = \vec{b}$. The two types of quantum circuits are denoted as $h_test$ and $specialh_test$ below. As in Figure \ref{fig:vqls}, both circuits have 9 parameters to optimize and uses 4 qubits.

\begin{figure}
	\includegraphics[width=0.6\textwidth]{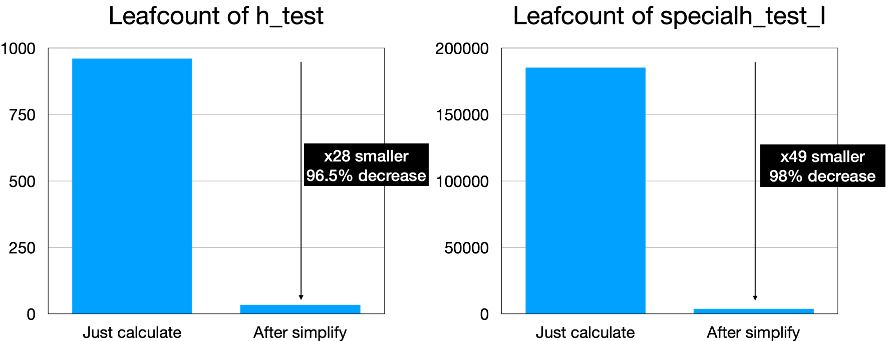}
	\centering
	\caption{Comparison of leafcount of the result of two circuits in VQLS }
	\label{fig:leafcount}
\end{figure}

Figure \ref{fig:leafcount} is the comparison of leafcount, or the expression size, of the result of two circuits VQLS. The expression size after simplification decreased 28 and 49 times less than before simplification and if we consider the circuit usage ratio, the overall size decreases 42 times smaller than the original method. Hence, one can expect 42 times speedup in VQLS learning time after simplification. To confirm this expectation, we realized the algorithm testing procedure as following.

\begin{figure}
	\includegraphics[width=0.6\textwidth]{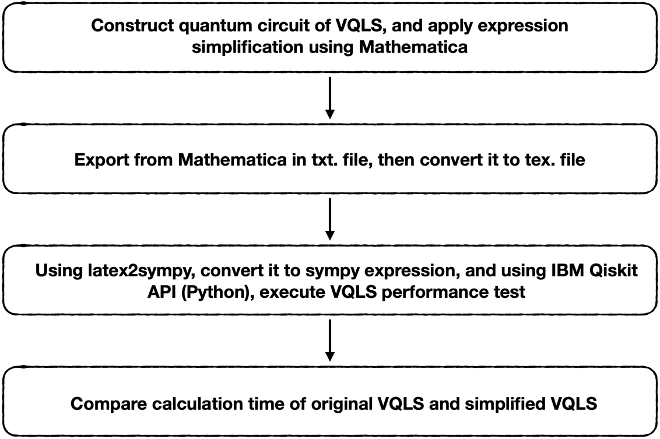}
	\centering
	\caption{We realized Measurement Simplification based VQLS by the following procedure. Calculation time can be estimated and can be compared with the original VQLS.}
	\label{fig:procedure}
\end{figure}

\begin{figure}
	\includegraphics[width=0.6\textwidth]{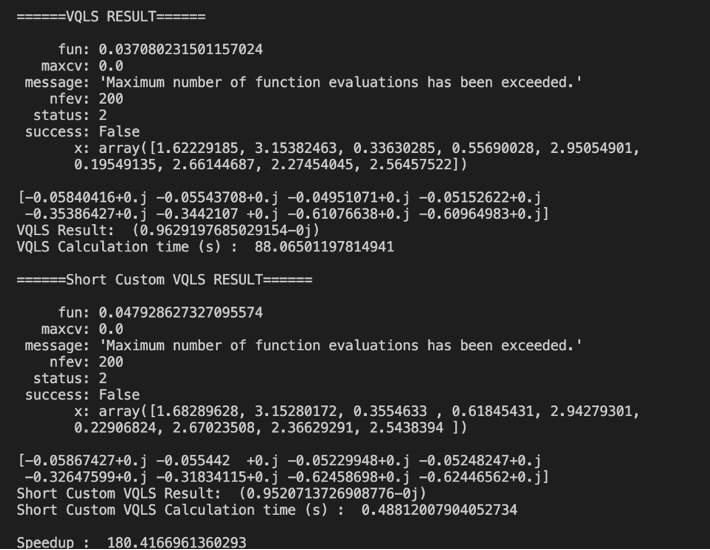}
	\centering
	\caption{Output display of the Measurement Simplification based VQLS. One can confirm the calculation time in original VQLS and simplification based VQLS. The ratio of two calculation times is defined as speedup.}
	\label{fig:output}
\end{figure}

Figure \ref{fig:procedure} is the output display of the python program we made for tesing the performance of Measurement Simplification based VQLS. We have executed the algorithm for number of qubits $n=3$, $A=0.45I+0.55Z_3$. The output shows each solution of linear equation using VQLS and simplification based VQLS, the elapsed time in second for each method, and the speedup factor, which is defined as the ratio of calculation time before and after simplification. The figure is reporting 180 times speedup after simplification.

\begin{figure}
	\includegraphics[width=0.6\textwidth]{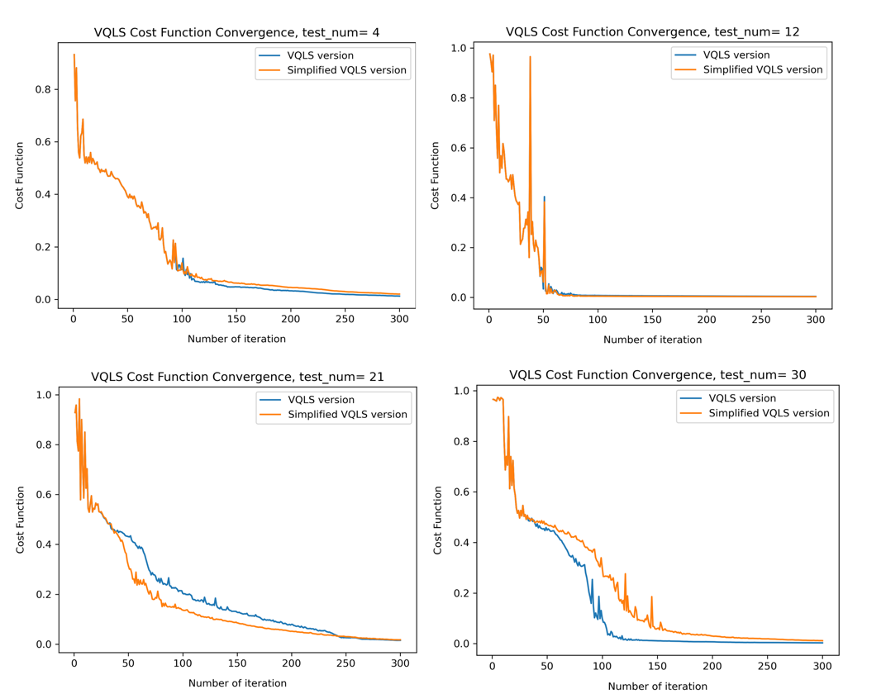}
	\centering
	\caption{Cost function w.r.t number of iteration graph. The original VQLS and simplification based VQLS shows similar convergence graph for most cases.}
	\label{fig:cost}
\end{figure}

Figure \ref{fig:cost} is showing the cost function convergence curve for changing the initial parameter value of VQC. All cases showed sufficient convergence of cost function to 0 in 200 iterations. $Test_num$ = 4, 12 graph shows the typical convergence curve which illustrates most of the cases and it implies VQLS and simplfication based VQLS fundamentally evaluates exact same value. Few exception cases, such as $Test_num$ = 21, 30, showed difference between the two convergence curves, and the error appears to be originated from the difference in the number of significant digits from the implementation method.

\begin{figure}
	\includegraphics[width=0.6\textwidth]{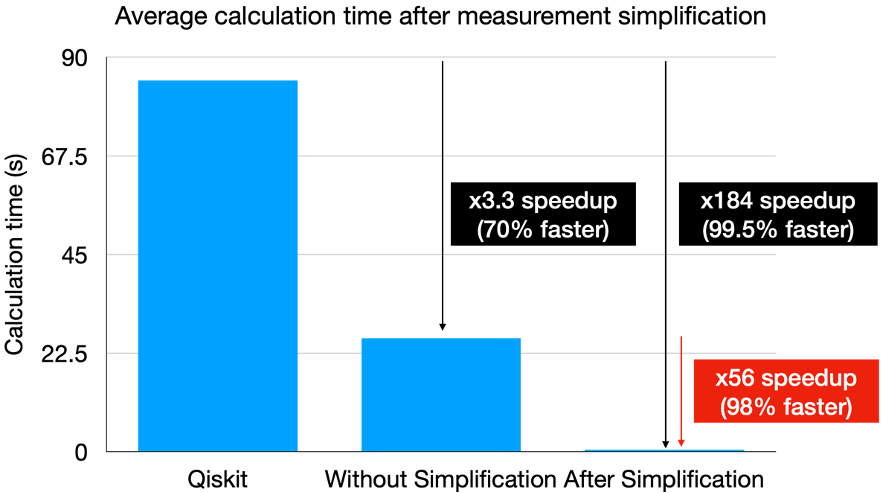}
	\centering
	\caption{Comparison of average calculation time after Measurement Simplification. The simplification reduced the calculation time 56 times shorter, and this is 98\% speedup.}
	\label{fig:simplify2}
\end{figure}

\begin{figure}
	\includegraphics[width=0.6\textwidth]{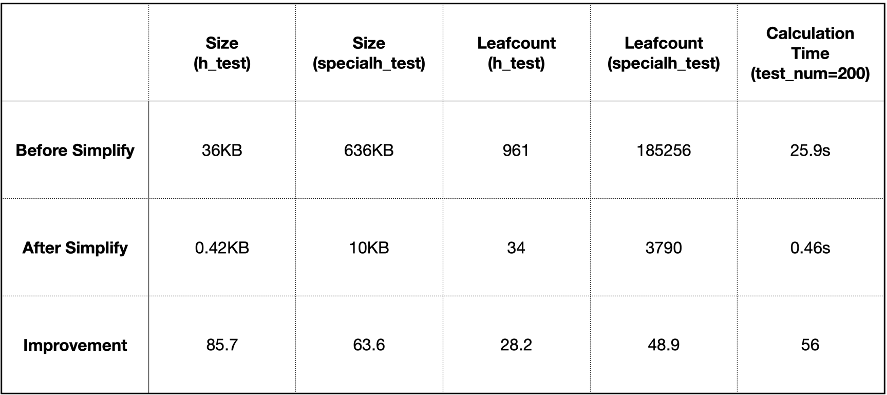}
	\centering
	\caption{Comparison of file size, expression size (leafcount), and calculation time for 200 iterations, of VQLS before and after Measurement Simplification. After simplification, file size has decreased 85.7, 63.6 times smaller for h\_test and specialh\_test circuit respectively. Expression size has decreased 28.2, 48.9 times smaller respectively and overall calculation time has decreased 56 times shorter.}
	\label{fig:table}
\end{figure}

Figure \ref{fig:simplify2}, \ref{fig:table} shows the effect of simplification at VQLS. Figure \ref{fig:simplify2} shows the comparison of calculation time between VQLS algorithm proposed from IBM Qiskit, our Mathematica, python based VQLS algorithm without simplification (S0) and with simplification (S1). We changed the initial parameter value with test set of dimension 100, and calculated the average value. We used COBYLA for the classical optimizer and for each test, and the number of iteration was fixed as 300.

First, there was 3.3 times speedup in calculation time comparing Qiskit and S0. We implemented our VQLS algorithm exploiting Lambdify function for substituting the parameter constants after the expression calculating process, and we interpreted this 3.3 times speedup coming from the modular function performance handling constant substitutions. Therefore, we interpreted the improvement coming from Measurement Simplification with respect to S0's calculation time.

Second, if we compare the result of Qiskit and S1, it shows 184 times speedup. This implies if the test set is large or the number of iteration is large, the simplification based VQLS's power will dramatically expose.

Third, comparing the result of S0 and S1, the ratio between the two times is the actual metric for estimating improvement derived from Measurement Simplification. The ratio between the calculation time base d on S0 to S1 was 56. This implies Simplification based VQLS was 56 times faster than the original VQLS method. This result is similar to the result from the previous leafcount analysis that showed 42 times improvement, and may conclude the improvement was based on expression simplification.

Figure \ref{fig:table} shows the comparison of file size, expression size (leafcount), and calculation time for 200 iterations, of VQLS before and after Measurement Simplification. After simplification, file size has decreased 85.7, 63.6 times smaller for h\_test and specialh\_test circuit respectively. Expression size has decreased 28.2, 48.9 times smaller respectively and overall calculation time has decreased 56 times shorter. Therefore, Measurement Simplification can also greatly decrease require memory capacity in addition to the decrease of calculation time.

\subsection{Application in Quantum Chemistry and Quantum Machine Learning (QML)}

\begin{figure}
	\includegraphics[width=0.6\textwidth]{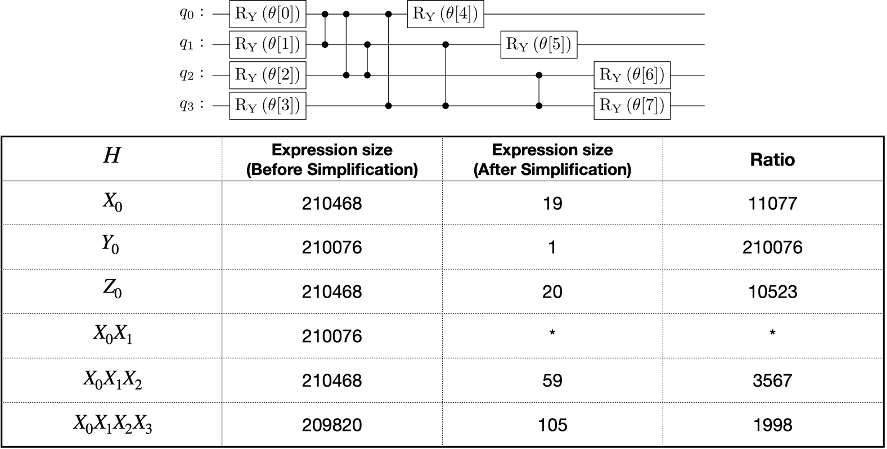}
	\centering
	\caption{Quantum Finance: Portfolio Optimization VQC and table showing the effect of simplification. After simplification, the expression size dramatically decreases.}
	\label{fig:portfolio}
\end{figure}

\begin{figure}
	\includegraphics[width=0.6\textwidth]{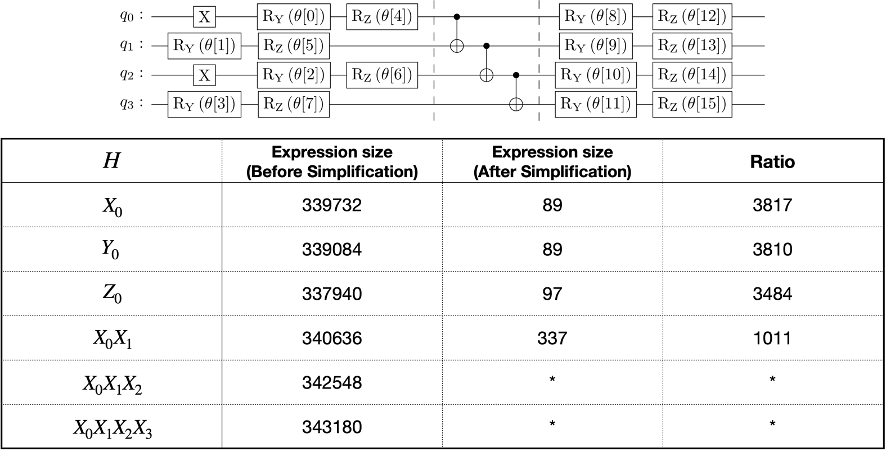}
	\centering
	\caption{Quantum Chemistry: Band gap calculation of OLED molecule VQC and table showing the effect of simplification. After simplification, the expression size dramatically decreases.}
	\label{fig:chemistry}
\end{figure}

\begin{figure}
	\includegraphics[width=0.6\textwidth]{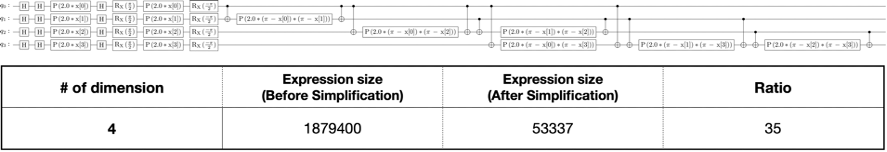}
	\centering
	\caption{ Quantum Machine Learning: Image Classification VQC and table showing the effect of simplification. After simplification, the expression size dramatically decreases.}
	\label{fig:image}
\end{figure}

2021 IBM Qiskit Hackathon presented 4 quantum algorithms that solve important problems in the field of Quantum Finance: Portfolio Optimization, Quantum Chemistry: Band gap calculation of OLED molecule, Quantum Machine Learning: Image Classification, Quantum Optimization: Battery Revenue Optimization. The main circuit of first 3 algorithms has a similar structure to the VQC we have analyzed previously. For example, in the case of Portfolio Optimization (Figure \ref{fig:portfolio}) and Quantum Chemistry (Figure \ref{fig:chemistry}) exploit the linear entangled VQC, and Image Classification (Figure \ref{fig:image} exploits the full entangled VQC. Noticing the equivalence of VQC structure, we applied Measurement Simplification to each algorithm, and investigated the expression size improvements.

Figure \ref{fig:portfolio} shows the part of main VQC used in Quantum Finance: Portfolio Optimization. Portfolio Optimization is the task of finding the optimal portfolio ratio that maximizes the expected profit and this can be computed by quantum computer by Variational Quantum Eigensolver(VQE). Specifically, by parametrized quantum circuit $U(\vec{\theta})$, evaluate the expectation value of Hamiltonian H which is corresponding to the expected profit and the classical optimizer gives the feedback to the quantum circuit in the direction to find the optimal parameter $\vec{\theta}$. The VQC depicted at Figure \ref{fig:portfolio} is the U circuit. For $\ket{\psi} = U\ket{0^n}$, the expectation value of Hamiltonian is $\bra{\psi}H\ket{\psi} = \bra{0^n}U^\dagger{}HU\ket{0^n}$, where the VQE's Hamiltoninan is defined as below.

\begin{equation}
    H = \sum_{i,\alpha} h_\alpha^i \sigma_\alpha^i + \sum_{i,j,\alpha,\beta} h_{\alpha, \beta}^{i,j} \sigma_\alpha^i \sigma_\beta^j + \cdots
\end{equation}

${h_\alpha^i}, {\sigma_\alpha^i}$ denote $ith$ qubit, $\alpha th$ spin state's coefficient and Pauli operator, respectively. Specifically, arbitrary Hamiltonian expressed by n qubits can be written as the linear combination of Pauli operators' tensor product, and each component can be readily implemented by quantum circuits. The VQE method is the method of calculating the each component's expectation value in quantum circuit and calculating the linear combination of the coefficients in classical computer. Therefore, for VQE, to calculate the expectation value $\bra{0^n}U^\dagger{}HU\ket{0^n}$, one may apply quantum circuits that is equivalent to $U^\dagger{} \sigma_\alpha^i U$ to $\ket{0^n}$ and just read the first amplitude from the output state vector, so it is reasonable to expect great improvement after Measurement Simplification. To verify this, we implemented VQC using Mathematica and simplified specific Hamiltonian components that correspond to the Portfolio Optimization problem. After simplification, as shown in Figure \ref{fig:portfolio}, improvement, or decrease of expression size, from 1000 times up to 10000 times was shown. The circuit was composed of 4 qubits and the number of parameters was 8. Item that took long time (more than  1 hour) in Mathematica simplify function was indicated as * mark. 

Figure \ref{fig:chemistry} shows the $U(\vec{\theta})$ circuit that is used in Quantum Chemistry: Band gap calculation of OLED molecule problem. It is also a VQE based algorithm, and when processed Measurement Simplification for same Hamiltonian components, the expression size was decreased 1000 times up to 3000 times. The number of qubits is 4 and the number of parameters used is 16.

Figure \ref{fig:image} shows the $U(\vec{\theta})$ circuit that is used in Quantum Machine Learning: Image Classification problem. The algorithm used Quantum Kernel based Quantum Support Vector Machine (QSVM) method to classify MNIST dataset and to reduce the required number of qubits, it used PCA to compress the dimension of the input data to 5. QSVM method utilizes quantum kernel that can be efficiently computed buy quantum circuit, and 
based on the measurement result of the quantum circuits, exploits classical SVM to classify the dataset. The quantum kernel is defined as below.

\begin{equation}
    K_ij=|\bra{0^n}U_{\Phi(x_j)}^\dagger{}U_{\Phi(x_i)}\ket{0^n}|^2
    =|\braket{\phi(x_j)^\dagger{}}{\phi(x_i)}|^2
\end{equation}

To verify the improvement due to the simplification, we chose the case for number of qubits = 4, U = Pauli feature map, and as the result, the expression size was decreased 35 times smaller. The number of parameters for the circuit is 8.


\section{Conclusion}
\label{sec:conclusion}

In this research, we have investigated when only partial information is required to run Variational Quantum Algorithm on a simulator, a processing named Measurement Simplification can reduce the expression size for evaluation, and that it is possible to gain great improvements in the aspect of computation speed and required memory capacity. The partial information of a state vector can refer to a specific qubit's measurement probability or an amplitude of a specific basis. Among the various VQAs, we have applied the simplification to quantum circuits exploited in QDRL, VQLS, VQE, Quantum kernel based QSVM, and depending to the structure of the circuit, expression size decreased from 3 times up to 10000 times. As Various VQAs are being actively discovered, Measurement Simplification can be applied to wide class of variational quantum circuits and algorithms.

To verify whether Measusrement Simplification actually affects the computation time, we have tested on the VQLS algorithm, and as a result, the expression size decreased 42 times smaller, and the calculation time decreased 56 times shorter. From the result, we were able to confirm that by the expression size decrease from the expression simplification derives the decrease of computation time. Hence, as the number of iteration is large, and as the dimension of dataset is large, the power of simplification would be large for executing VQAs.

The variational circuits that is utilized at Quantum Finance, Quantum Chemistry, Quantum Machine Learning problems were the case where only a single amplitude is required, and in these cases, the decrease of expression size after simplification was dramatically large. The decrease ratio was exhibited from 1000 to 10000, and based on the analysis, future work can realized the simplified version algorithms for such case and verity the actual decrease in computation time and memory usage.

Further work is needed to investigate the condition about when considerable simplification is achieved. Each VQC structures exhibit different improvement factor after simplification, and if the condition about when strong simplification is achieved, efficient application of Measurment Simplifcation would be possible. There exists some cases when simplification based on Mathematica was slow, and if the condition is thoroughly understood, such problems can also be resolved.

\section*{Acknowledgments}
This Project was financed by the Samsung Electronics "Beyond Limit" project.

\bibliographystyle{unsrt}  
\bibliography{references}  

\end{document}